\documentstyle[12pt]{article}
\def\beq{\begin{equation}}
\def\eeq{\end{equation}}
\def\bea{\begin{eqnarray}}
\def\eea{\end{eqnarray}}
\def\bq{\begin{quote}}
\def\eq{\end{quote}}
\def\ve{\vert}
\def\nnb{\nonumber}
\def\ga{\left(}
\def\dr{\right)}
\def\aga{\left\{}
\def\adr{\right\}}

\def\rar{\rightarrow}
\def\nnb{\nonumber}
\def\la{\langle}
\def\ra{\rangle}
\def\nin{\noindent}
\begin{document}
\topmargin -1.5cm
\oddsidemargin +0.2cm
\evensidemargin -1.0cm
\pagestyle{empty}
\begin{flushright}
{CERN-TH.7277/94}\\
PM 94/11
\end{flushright}
\vspace*{5mm}
\begin{center}
\section*{\bf
HEAVY QUARKS FROM \\
QCD SPECTRAL SUM RULES$^{*)}$}
\vspace*{0.5cm}
{\bf S. Narison} \\
\vspace{0.3cm}
Theoretical Physics Division, CERN\\
CH - 1211 Geneva 23, Switzerland\\
and\\
Laboratoire de Physique Math\'ematique\\
Universit\'e de Montpellier II\\
Place Eug\`ene Bataillon\\
34095 - Montpellier Cedex 05, France\\
\vspace*{1.5cm}
{\bf Abstract} \\ \end{center}
\vspace*{2mm}
\noindent
This is a short review of the present status of the dynamics of the
heavy quarks extracted from QCD spectral sum rules.
We discuss the determination of the ``perturbative"
quark mass, the decay constants of the $D$ and $B$ mesons, the
form factors of the semileptonic and radiative decays of the
 $B$ mesons, the slope of the Isgur--Wise function
and the emerged value of $V_{cb}$. We mention shortly some
properties of the hybrid  and $B_c$ mesons.

\vspace*{2.0cm}
\noindent
\rule[.1in]{16.0cm}{.002in}

\noindent
$^{*)}$ Talk given at the {\it XXIXth Rencontre de Moriond on ``
QCD and High Energy Hadronic Interactions"}, M\'eribel, Haute-Savoie,
19--26th March 1994 and at the {\it
Rencontre entre experimentateurs et theoriciens
sur ``la violation de CP"}, Orsay, 6--7th May 1994.
\vspace*{0.5cm}
\noindent


\begin{flushleft}
CERN-TH.7277/94 \\
PM 94/11\\
\today
\end{flushleft}
\vfill\eject
\pagestyle{empty}

 \setcounter{page}{1}
 \pagestyle{plain}
\section{Introduction} \par
We have been
 living with QCD spectral sum rules (QSSR) (or QCD sum rules, or
ITEP sum rules, or hadronic sum rules...)
for 15 years, within the
impressive ability
of the method for describing the complex phenomena of hadronic
physics with the few universal ``fundamental" parameters of the QCD
Lagrangian
(QCD coupling $\alpha_s$, quark masses
and  vacuum condensates built from the quarks
and/or gluon fields, which parametrize the
non-perturbative phenomena). The approach might be very close to the
lattice calculations as it also uses the first principles of QCD, but
unlike the case of the lattice, which is based  on
sophisticated numerical simulations, QSSR is
quite simple as it is a semi-analytic approach based
on a semiperturbative expansion and Feynmann graph
techniques implemented in an Operator Product Expansion (OPE),
where  the condensates contribute as higher dimension
operators in the OPE. In this approach, one
can {\it really control and in some sense understand} the origin
of the numbers obtained from the analysis.

\nin
Since its early days and up to now, the field has,
unfortunately, suffered from some apparent controversy and (worse) from
some irrationnal emotional fights between different groups, which
obscure its beauty. But, when you
start {\it d'y fourrer votre nez }, you realize that the QCD spectral
sum rule is one of the most interesting discoveries
 of the last decade since,
with its simplicity, it can describe in an elegant way the complexity
of the hadron phenomena,
without waiting for a complete understanding of the confinement problem.

\nin
One can fairly say that QCD spectral sum rules already started, before
QCD, in
the time of current algebra in 60,
when peoples proposed different {\it ad hoc} superconvergence
sum rules, especially the Weinberg and Das--Mathur--Okubo
sum rules, which came under
control only within the advent of QCD \cite{FLO}. However, the main flow
comes from the classic paper of Shifman--Vainshtein--Zakharov
\cite{SVZ} (hereafter referred to as SVZ), which goes beyond
the {\it na\"\i ve} perturbation
theory thanks to the inclusion of the vacuum condensate effects in
the OPE. More
details and more complete discussions of QSSR and its various
applications to hadron physics can be found, for instance, in
\cite{SNB}.

\nin
In this talk, I shall present aspects of QSSR in the analysis of the
properties of heavy flavours. As I am limited in space-time
(an extended and updated version of this talk will be
published in the Proceedings of QCD 94 at Montpellier), I cannot
cover in detail all QSSR applications
to the heavy-quark physics here. I will only
shortly discuss the following topics,
which I think are important in the development of the understanding
of the heavy-quark properties. These concern the determination
of the:

\nin
-- heavy-quark ``perturbative" pole and running masses,

\nin
-- decay constants of the
$D$, $B$ and $D_s$, $B_s$ mesons and the $B_B$-parameter,

\nin
-- semileptonic and radiative decay form factors of the $B$,

\nin
-- slope of the Isgur--Wise function and the emerged
value of $V_{cb}$,

\nin
-- properties of hybrids and $B_c$ mesons.

 \section{ QCD spectral sum rules}
In order to illustrate the QSSR method in a pedagogical way, let us consider
the two-point correlator:
\bea
\Pi^{\mu\nu}_b(q^2,M^2_b) &\equiv& i \int d^4x ~e^{iqx} \
\la 0\vert {\cal T}
J^{\mu}_b(x)
\ga J^{\nu}_b(o)\dr ^\dagger \vert 0 \ra \\ \nnb
&=& -\ga g^{\mu\nu}q^2-q^\mu q^\nu \dr \Pi_b(q^2,M^2_b),
\eea
where $J^{\mu}_b(x) \equiv \bar b \gamma^\mu b (x)$ is the local vector
current of the $b-$quark. The correlator obeys the well-known
K\"allen-Lehmann dispersion relation:
\beq
\Pi_b(q^2,M^2_b) = \int_{4M^2_b}^{\infty} \frac{dt}{t-q^2-i\epsilon}
{}~\frac{1}{\pi}~\mbox{Im}  \Pi_b(t) \ \ \ + \ \ \ \mbox{subtractions},
\eeq
which expresses in a clear way the {\it duality} between the spectral
function Im$ \ \Pi_b(t)$, which can be measured experimentally, as here
it is related to the $e^+e^-$ into $\Upsilon$-like states total
cross-section, while $\Pi_b(q^2,M^2_b)$ can be calculated directly in
QCD, even at $q^2=0$,
thanks to the fact that $M^2_b-q^2 \gg \Lambda^2$.
The QSSR is an improvement on the previous
dispersion relation.

\nin
In the QCD side, such an improvement is achieved by adding
to the usual perturbative expression of the correlator,
the non-perturbative contributions as parametrized by the vacuum
condensates of higher and higher dimensions in the OPE \cite{SVZ}:
\beq
\Pi_b(q^2,M^2_b) \simeq \sum_{D=0,2,4,...}\frac{1}{\ga M^2_b-q^2 \dr^{D/2}}
\sum_{dim O=D} C^{(J)}(q^2,M^2_b,\mu)\la O(\mu)\ra,
\eeq
where $\mu$ is an arbitrary scale that separates the long- and
short-distance dynamics; $C^{(J)}$ are the Wilson coefficients calculable
in perturbative QCD by means of Feynman diagrams techniques:
$D=0$
corresponds to the case of na\"\i ve perturbative contribution;
$\la O \ra$ are
the non-perturbative condensates built
 from the quarks or/and gluon
fields. For $D=4$, the condensates that can be formed are the
quark $M_i \la\bar \psi \psi \ra$ and gluon $\la\alpha_s G^2 \ra$
ones; for
$D=5$, one can have the mixed quark-gluon condensate $\la\bar \psi \sigma_
{\mu\nu}\lambda^a/2 G^{\mu\nu}_a \psi \ra$, while for $D=6$ one has, for
instance, the
triple gluon $gf_{abc}\la G^aG^bG^c \ra$ and the four-quark
$\alpha_s \la \bar \psi \Gamma_1 \psi \bar \psi \Gamma_2 \psi \ra$, where
$\Gamma_i$ are generic notations for any Dirac and colour matrices. The
validity of this expansion has been understood formally using renormalon
techniques \cite{MUELLER} and by building  renormalization invariant
combinations of the condensates (Appendix of \cite{PICH}).
The SVZ-expansion is
phenomenologically confirmed
from the unexpected
accurate determination of the QCD coupling $\alpha_s$ from semi-inclusive
tau decays \cite{PICH, ALFA}.
In the present case of heavy-heavy correlator
the OPE is much simpler, as one can show \cite{HEAVY, BC}
that the heavy-quark condensate
effects can be included into those of the gluon condensates, such
that up to $D\leq 6$, only the $G^2$ and $G^3$ condensates appear in
the OPE. Indeed, SVZ has, originally, exploited this feature for their
first estimate of the gluon condensate value.

\nin
For the phenomenological side, the improvement comes from the uses of
either finite number of derivatives and finite values of $q^2$
(moment sum rules):
\beq
{\cal M}^{(n)} \equiv \frac{1}{n!}\frac{\partial^n \Pi_b(q^2)}
{\ga \partial q^2\dr^n} \Bigg{\vert} _{q^2=0}
= \int_{4M^2_b}^{\infty} \frac{dt}{t^{n+1}}
{}~\frac{1}{\pi}~ \mbox{Im}  \Pi_b(t),
\eeq
or infinite number of derivatives and infinite values of $q^2$, but
keeping their ratio fixed as $\tau \equiv n/q^2$
(Laplace or Borel or exponential sum rules):
\beq
{\cal L}(\tau,M^2_b)
= \int_{4M^2_b}^{\infty} {dt}~\mbox{exp}(-t\tau)
{}~\frac{1}{\pi}~\mbox{Im} \Pi_b(t).
\eeq
There also exists non-relativistic versions of these two sum rules,
which are convenient quantities to work with in the large-quark-mass
limit. In these cases, one introduces non-relativistic
variables $E$ and $\tau_N$:
\beq
t \equiv (E+M_b)^2 \ \ \ \ \mbox{and} \ \ \ \  \tau_N = 4M_b\tau .
\eeq
In the previous sum rules,
the gain comes from the weight factors, which enhance the
contribution of the lowest ground state meson to the spectral integral.
This fact makes the simple duality ansatz parametrization:
\beq
``\mbox{one resonance}"\delta(t-M^2_R)
 \ + \ ``\mbox{QCD continuum}" \Theta (t-t_c),
\eeq
of the spectral function
to give a very good description of the spectral integral, where the
resonance enters via its coupling to the quark current. In the case
of the $\Upsilon$, this coupling can be defined as:
\beq
\la 0\vert \bar b\gamma^\mu b \vert \Upsilon \ra =
 \sqrt{2} \frac{M^2_\Upsilon}
{2\gamma_\Upsilon}.
\eeq
The previous
feature
has been tested in the light-quark channel from the $e^+e^- \rar$
I=1 hadron data and in the heavy-quark ones from the
$e^+e^- \rar \Upsilon$ or $\psi$ data, within a good
accuracy.
To the previous sum rules, one can also add the ratios:
\beq
{\cal R}^{(n)} \equiv \frac{{\cal M}^{(n)}}{{\cal M}^{(n+1)}}~~~~~~~
\mbox{and}~~~~~~~
{\cal R}_\tau \equiv -\frac{d}{d\tau} \log {{\cal L}},
\eeq
and their finite energy sum rule (FESR) variants, in order to fix
the squared mass of the ground state.
\nin
In principle, the pairs $(n,t_c)$, $(\tau,t_c)$ are free external
parameters in the analysis, so that the optimal result should be
insensitive to their variations. Stability criteria, which are equivalent
to the variational method, state that the optimal results should
be obtained at the minimas or at the inflexion points in $n$ or $\tau$,
while stability in $t_c$ is useful to control the sensitivity of the
result in the changes of $t_c$-values. To these stability criteria are
added constraints from local duality FESR, which
correlate the $t_c$-value to those of the ground state mass and
coupling \cite{FESR}. Stability criteria have also been tested in
models such as
harmonic oscillator, where the exact and approximate
solutions are known \cite{BELL}. The {\it most conservative
optimization criteria} which include various types of optimizations
in the literature are the following: the
optimal result is obtained in the region,
from the beginning of $\tau / n$ stability (this corresponds in most
of the cases to the so-called plateau discussed often in the literature,
but to my opinion, the interpretation of this nice plateau as a good
sign of a good continuum model is not sufficient, in the sense
that the flatness of the
 curve extends in the uninteresting high-energy region where the
properties of the ground state are lost),
until the beginning of the $t_c$
stability, where the value of $t_c$ corresponds to about the one fixed by
FESR duality constraints.
The earlier {\it sum rule window} introduced by SVZ, stating that the
optimal result should be in the region where both the non-perturbative
and continuum contributions are {\it small} is included in the previous
region.
 Indeed, at the stability
point, we have an equilibrium between the continuum and non-perturbative
contributions, which are both small,
while the OPE is still convergent  at this point.
\section{The ``perturbative" masses}
We shall only discuss the determinations of the quark
masses from relativistic sum rules which use a truncated perturbative
series, where
the $perturbative$ masses entering
 in the sum rule analysis are then $well$-$defined$. This is not
the case of non-relatisvistic sum rules and heavy-quark
 effective theory where the quark is considered to be infinitely massive
and the QCD series is summed up. In this case,
the pole mass is $ill$-$defined$,
due to the parasitic renormalon effects, which induce after a
resummation of the QCD series at large order a ``non-perturbative"
piece of the order of $\Lambda$ (however, this effect is
regularization-scheme dependent and is absent in the $\overline {MS}$
-scheme)
\cite{BENEKE}. Also, in this infinite
mass limit, the effects of the coulombic terms are important which render
the estimate from non-relativistic sum rules very sensitive to the value
of $\alpha_s$ used in the analysis. These $non$-$perturbative$ effects
induce a kind of $dressed$ quark where the associated $ill$-$defined$
pole mass has a
strength similar to the one used in potential models, but
higher than the one obtained from the standard relativistic
sum rule at finite values of the quark mass and for a truncated
QCD series. In the same way,
non-perturbative effects, due to hadronization,
shift, to higher values, the mass
obtained from a fit of the inclusive decays and deep
inelastic scattering data. Within the same reasoning, we would also
expect that the pole mass from the lattice contains a
non-perturbative piece.

\nin
The masses obtained from the relativistic sum rules should be identified,
like in the case of the light quarks,
with the so-called current mass entering in the QCD Lagrangian, which
is a $well$-$defined$ quantity in perturbation theory within the
$\overline {MS}$-scheme.
Using relativistic sum rules in the
$\psi$ and $\Upsilon$ channels, different groups have obtained
the Euclidean mass \cite{SNB}:
\beq
M_c(p^2=-M^2_c)  \simeq  (1.26 \pm 0.02)~ \mbox{GeV} ~~~~~~~
M_b(p^2=-M^2_b)  \simeq  (4.18 \pm 0.02)~ \mbox{GeV},
\eeq
defined in the Landau gauge\footnote[1]{We have included in the
average the
slightly lower and more precise value
$M_b(p^2=-M^2_b)=( 4.17 \pm 0.02)$ GeV \cite{REINDERS} obtained in
this channel from relativistic moment sum rules.}.
 However, the use of this
gauge-dependent mass is not inconvenient as we know
that the correlator entering
the sum rule analysis is gauge-invariant. Instead, the use of the
Euclidean mass
minimizes the size of the radiative corrections at the order where the
sum rule analysis is done, making
the result insensitive to the error in the value of
$\alpha_s$. One can translate this result into
 the  $perturbative$ pole mass through the relation:
\beq
M_Q (p^2=M^2_Q) \simeq
M_Q(p^2=-M^2_Q)\ga 1+ 2\log 2 \frac{\alpha_s}{\pi}\dr ,
\eeq
 from which one can deduce the value of the
$perturbative~ pole$ masses which are gauge-invariant:
\beq
M_c(p^2=M^2_c)  \simeq  (1.46 \pm 0.05)~ \mbox{GeV} ~~~~~~~
M_b(p^2=M^2_b)  \simeq  (4.64 \pm 0.06) ~\mbox{GeV},
\eeq
for $\Lambda_5 \simeq (180 \pm 80)$ MeV.
 One can also estimate the
$perturbative~b$-pole mass by
using the ratios of the relativistic
 sum rules in the $B$ and $B^*$ channels. One obtains \cite{SNB}:
\beq
M_b(p^2=M^2_b) \simeq (4.56 \pm 0.05)~ \mbox{GeV}.
\eeq
By taking
the average of their values from the previous two independent sources,
one can deduce (see also \cite{SN1}):
\beq
M_b(p^2= M^2_b) \simeq (4.59 \pm 0.04)~ \mbox{GeV}.
\eeq
The value of the $perturbative~b$-pole mass is definitely lower
than the ones, in the range 4.8-5.3 GeV, from non-relativistic
 approaches, inclusive data and lattice calculations,
  due to the presence of the
$non$-$perturbative$ components induced by the summation of the
QCD series
in these latter cases.
\nin
The running masses in the $\overline {MS}$-scheme can be deduced
directly from the Euclidian masses in (10)
 or equivalently
from the $perturbative~ pole$ masses in (12) and (14)
through (derived for the
first time in \cite{SN1}):
\beq
M_Q (p^2=M^2_Q) \simeq
\overline{M}_Q
(M_Q^2)\ga 1+ \frac{4}{3}\frac{\alpha_s}{\pi}\dr.
\eeq
One obtains:
\beq
\overline{M}_c(1 \ \mbox{GeV}) \simeq (1.40 \pm 0.06)~\mbox{ GeV} ~~~~~~~
\overline{M}_b(1 \ \mbox{GeV}) \simeq (5.87 \pm 0.06)~\mbox{ GeV},
\eeq
which, combined with the $s$-quark mass values in \cite{SN1}, gives:
\beq
M_b/m_s \simeq 36.7 \pm 2.2,
\eeq
a result of a great interest for model-building and GUT-phenomenology.
\section{The pseudoscalar decay constants and $B_B$}
The decay constants $f_P$ of a pseudoscalar meson $P$ are defined as:
\beq
(m_q+M_Q)\la 0\vert \bar q (i\gamma_5)Q \vert P\ra
 \equiv \sqrt{2} M^2_P f_P,
\eeq
where in this normalisation $f_\pi = 93.3$ MeV.
A {\it rigorous}
upper bound on these couplings can be derived from the
second-lowest superconvergent moment:
\beq
{\cal M}^{(2)} \equiv \frac{1}{2!}\frac{\partial^2 \Psi_5(q^2)}
{\ga \partial q^2\dr^2} \Bigg{\vert} _{q^2=0},
\eeq
where $\Psi_5$ is the two-point correlator associated to the pseudoscalar
current. Using the positivity of the higher-state contributions to the
spectral function, on can deduce \cite{SNZ}:
\beq
f_P \leq \frac{M_P}{4\pi} \aga 1+ 3 \frac{m_q}{M_Q}+
0.751 \bar{\alpha}_s+... \adr,
\eeq
where one should not misinterpret the mass-dependence in this
expression compared to the one expected from heavy quark symmetry.
Applying this result to the $D$-meson, one obtains:
\beq
f_D \leq 2.14 f_\pi .
\eeq
Although
presumably quite weak, this bound, when combined with the recent
determination to two loops \cite{SN2}:
\beq
\frac{f_{D_s}}{f_D} \simeq (1.15 \pm 0.04)f_\pi ,
\eeq
implies
\beq
f_{D_s} \leq (2.46 \pm 0.09)f_\pi ,
\eeq
which is useful for a comparison with the recent measurement of $f_{D_s}$
from WA75: $f_{D_s} \simeq (1.76 \pm 0.52)f_\pi$ and from CLEO:
$f_{D_s} \simeq (2.61 \pm 0.49)f_\pi$.
One cannot push, however, the uses
of the moments to higher $n$-values in this $D$-channel, in order to
minimize the continuum contribution to the sum rule with the aim to
derive an estimate of the decay constant because the QCD-series
will not converge at higher $n$-values. The estimate of \cite{DOM}, based
on the lowest moment, indeed
suffers from the continuum sensitivity as
 a little change in the continuum threshold makes
a big change of the estimate, in such a way that the result becomes
unreliable. In the $D$-channel, the most appropriate sum rule is the
Laplace sum rule. The results from different groups are consistent
for a given value of the $c$-quark mass and
lead to the average \cite{SN3,SNA}:
\beq
f_D \simeq (1.31 \pm 0.12)f_\pi~~~~~~~ \Rightarrow ~~~~~~~~~
f_{D_s} \simeq (1.51 \pm 0.15)f_\pi .
\eeq
For the $B$-meson, one can either work with the Laplace, moments or
their non-relativistic variants. Given the previous value of $M_b$, these
different methods give consistent values of $f_B$, though the one
from the non-relativistic sum rule is very inaccurate due to the huge
effect of the radiative corrections in this method. The average value
of $f_B$ is \cite{SN4}:
\beq
f_B \simeq (1.60\pm 0.26)f_\pi ,
\eeq
while \cite{SN2}:
\beq
 \frac{f_{B_s }}{f_B}\simeq 1.16 \pm 0.04,
\eeq
where the most accurate estimate comes from the ``relativistic" Laplace
sum rule. One could notice, since the {\it first} result
$f_B \simeq f_D$ of \cite{SN3}, a large violation of the
scaling law expected from heavy-quark symmetry. Indeed,
 this is due to the large 1/$M_b$-correction
 found from the HQET sum rule \cite{BALL, NEU} and from the one in full
QCD \cite{SN4,SNA}:
\beq
f_B \sqrt{M_b} \simeq (0.42 \pm 0.07)~\mbox{GeV}^{3/2}\aga 1-\frac{
(0.88\pm
0.18)~\mbox{GeV}}{M_b}\adr,
\eeq
which is due to the meson-mass gap $\delta M \equiv M_B-M_b$ \cite{NEU}
and to the continuum energy $E_c$ \cite{SN4, ZAL} ($E_c \simeq
\frac{3}{4} \delta M $ \cite{SNA}):
\beq
f_B \sqrt{M_b} \simeq \frac{1}{\pi}E_c^{3/2}\aga 1-\frac{\delta M}{M_b}
-\frac{3}{2}\frac{E_c}{M_b}+...\adr.
\eeq
One can  notice
that the apparent
disagreement among different existing
QSSR numerical results in the literature is
mainly due to the
different values of the quark masses used because the decay constants are very
sensitive to that quantity as shown explicitly in \cite{SN2}. Indeed,
one can also
exploit this sensitivity in order to deduce the value of the quark mass
when the experimental measurement of these decay constants will
become available.

\nin
Finally, let me also mention that we have also tested the validity
of the vacuum saturation $B_B=1$ of the bag parameter, using a
sum rule analysis of the four-quark two-point correlator to two loops
\cite{PIVO}. We
found that the radiative corrections are quite small. Under some
 physically reasonnable assumptions for the spectral function, we found
that the vacuum saturation estimate is only violated by about $15\%$,
giving:
\beq
 B_B \simeq 1 \pm 0.15.
\eeq
By combining this result with the one for $f_B$ in (25), we deduce:
\beq
f_B\sqrt{B_B} \simeq  211 \pm 36~{\mbox MeV},
\eeq
where we have used the normalization $f_\pi= 132$ MeV, which is $\sqrt 2$
the one defined in (18).
These previous results are in excellent agreement with the
present lattice calculations \cite{PENE}. In particular:
\beq
f_B\sqrt{B_B} \simeq  220 \pm 40~{\mbox MeV},
\eeq
A tentative average of these two numbers leads to the ``world average":
\beq
f_B\sqrt{B_B} \simeq  215 \pm 27~{\mbox MeV},
\eeq
which is useful in the phenomenological analysis of the
$B^0$-$\bar{B^0}$-mixing.
\section{Rare and semileptonic decays}
One can extend the analysis done for the two-point correlator to the
more complicated case of three-point function in order to study the form
factors related to the $B\rar K^*\gamma$ and $B\rar \rho/\pi$ semileptonic
decays. In so doing, one can consider the generic three-point function:
\beq
V(p,p',q^2)\equiv -\int d^4x~ d^4y ~e^{i(p'x-py)} ~\la 0|{\cal T}
 J_L(x)O(0)J^{\dagger}_B(y)|0\ra
\eeq
where $J_L, ~J_B$ are the currents of the light and $B$-mesons; $O$
 is the
weak operator specific for each process (penguin for the $K^* \gamma$,
weak current for the semileptonic); $q \equiv p-p'$.
The vertex obeys the double dispersion
relation :
\beq
V(p^2,p'^2,q^2) \simeq \int_{M_b^2}^{\infty} \frac{ds}{s-p^2-i\epsilon}
\int_{m_L^2}^{\infty} \frac{ds'}{s'-p'^2-i\epsilon}
{}~\frac{1}{\pi^2}~ \mbox{Im}V(s,s',q^2)+...
\eeq
As usual, the QCD part enters in the LHS of the sum rule, while the
experimental observables can be introduced through the spectral function
after the introduction of the intermediate states. The improvement of the
dispersion relation can be done in the way discussed
previously for the two-point
function. In the case of the heavy-to-light transition,
 the only possible
improvement whith a good $M_b$-behaviour
at large $M_b$ is the so-called
hybrid sum rule (HSR) corresponding to the uses of
the moments for the heavy-quark
channel and to the Laplace for the light one \cite{SNA, SN5}:
\beq
{\cal H} (n, \tau') =\frac{1}{\pi^2} \int_{M^2_b}^\infty \frac{ds}{s^{n+1}}
\int_0^\infty ds'~e^{-\tau' s'}~\mbox{Im}V(s,s',q^2).
\eeq
We have
studied analytically the different form factors entering in the previous
processes \cite{SN6}. They are defined as:
\bea
\la\rho(p')\ve \bar u \gamma_\mu (1-\gamma_5) b \ve B(p)\ra
&=&(M_B+M_\rho)A_1
\epsilon^*_\mu -\frac{A_2}{M_B+M_\rho}\epsilon^*p'(p+p')_\mu \nnb \\
&&+\frac{2V}{M_B+M_\rho} \epsilon_{\mu \nu \rho \sigma}p^\rho p'^\sigma ,
\nnb \\
\la\pi(p')\ve
\bar u\gamma_\mu b\ve B(p)\ra &=& f_+(p+p')_{\mu}+f_-(p-p')_\mu , \nnb \\
\la \rho(p')\ve \bar s \sigma_{\mu \nu}\ga
\frac{1+\gamma_5}{2}\dr q^\nu b\ve B(p)\ra &=&
i\epsilon_{\mu \nu \rho \sigma}\epsilon^{*\nu}p^\rho p'^\sigma
F^{B\rar\rho}_1
\nnb \\
&&+ \aga \epsilon^*_\mu(M^2_B-M^2_{\rho})-\epsilon^*q(p+p')_{\mu}
\adr \frac{F^{B\rar \rho}_1}{2}.
\eea
We found that they are dominated, for $M_b \rar \infty$, by the
effect of the light-quark condensate, which dictates the $M_b$-behaviour
of the form factors to be
typically of the form:
\beq
F(0) \sim \frac{\la \bar dd \ra}{f_B}\aga 1+\frac{{\cal I}_F}
{M^2_b}\adr,
\eeq
where ${\cal I}_F$ is the integral from the perturbative triangle
graph, which is constant as $t'^2_cE_c/\la \bar dd \ra$ ($t'_c$ and
$E_c$ are the continuum thresholds of the light and $b$ quarks)
for large value of $M_b$. It
indicates that at $q^2=0$, all form factors behave like $\sqrt{M_b}$,
although, in most cases, the coefficient of the $1/M^2_b$ term is large.
The
study of the $q^2$-behaviours of the form factors indicates
 that with the
exception of the $A_1$-form factor, the $q^2$-dependence of the others
is only due to the non-leading (in $1/M_b$) perturbative graph such that
for $M_b \rar \infty$,
these  form factors remain constant from $q^2=0$ to $q^2_{max}$.
The resulting $M_b$-behaviour at $q^2_{max}$ is the one expected from the
heavy quark symmetry. The numerical
effect of this $q^2$-dependence at finite values of $M_b$ is a polynomial in
$q^2$, which mimics  quite well the usual pole parametrization for a pole mass
of about 6--7 GeV. The situation for the $A_1$ is drastically different from
the other ones as here the Wilson coefficient of the $\la \bar dd \ra$
condensate contains a $q^2$-dependence and reads:
\beq
A_1(q^2) \sim \frac{\la \bar dd \ra}{f_B}\aga 1-\frac{q^2}{M^2_b}
\adr ,
\eeq
which, for $q^2_{max} \equiv (M_B-M_\rho)^2$, gives the expected behaviour:
\beq
A_1(q^2_{max}) \sim \frac{1}{\sqrt{M_b}}.
\eeq
One can notice that the $q^2$-dependence of $A_1$ is in complete contradiction
with the pole behaviour, as has been also noticed in the numerical
analysis of \cite{BALL2}. It
is urgent and important to test this feature experimentally.
One can finally notice that due to the overall $1/f_B$ factor, all
form factors have a large $1/M_b$-correction.

\nin
In the numerical analysis, we obtain at $q^2=0$, the value of the $B\rar
K^*\gamma$ form factor:
\beq
F_1^{B\rar \rho } \simeq 0.27 \pm 0.03,~~~~~~~~~
\frac{F_1^{B\rar K^*}}{F_1^{B\rar \rho}}\simeq 1.14 \pm 0.02,
\eeq
which leads to the branching ratio $(4.5\pm 1.1)\times 10^{-5}$, in perfect
agreement with the CLEO data and with the estimate in \cite{ALI}.
One should also notice that, in this case,
the coefficient of the $1/M^2_b$ correction is very large,
which makes the extrapolation of the
the $c$-quark results to higher values of the
quark mass dangerous. This extrapolation is often done in some
lattice calculations.

\nin
For the semileptonic decays, QSSR
give a good determination of the ratios of
the form factors with the values \cite{SN5}:
\bea
&&\frac{A_2(0)}{A_1(0)} \simeq \frac{V(0)}{A_1(0)}
\simeq 1.11 \pm 0.01    \nnb  \\
&&\frac{A_1(0)}{F_1^{b\rar \rho}(0)} \simeq 1.18 \pm 0.06    \nnb \\
\eea
Combining these results with the ``world average" value of $f_+(0)=
0.25 \pm 0.02$ and the one of $F_1^{B\rar \rho}(0)$, one can deduce the
rate and polarization:
\bea
&&\Gamma_\pi \simeq (4.3\pm 0.7)
|V_{ub}|^2 \times 10^{12}~\mbox{s}^{-1} \nnb \\
&&\Gamma_\rho /\Gamma_\pi \simeq 0.9 \pm 0.2    \nnb \\
&&\Gamma_+ /\Gamma_- \simeq 0.20\pm 0.01   \nnb \\
&&\alpha \equiv 2\frac{\Gamma_L}{\Gamma_T}-1 \simeq -(0.60 \pm 0.01).
\eea
These results are much more precise than
the ones from a direct estimate of the absolute values of the form
factors due to the cancellation of systematic errors in the ratios. They
indicate that, we are on the way to reach $V_{ub}$ with a good accuracy
from the exclusive modes.
Also
here, due mainly to the non-pole behaviour of $A^B_1$,
the ratio between the widths into $\rho$ and into
$\pi$ is about 1, while in different pole models,
 it ranges from 3 to 10.
For the
asymmetry, one obtains a large negative value of $\alpha$, contrary to
the case of the pole models.
\section{Slope of the Isgur--Wise function and $V_{cb}$}
Let me now
discuss  the slope of the Isgur--Wise function. Taron--de Rafael
\cite{TARON}has exploited the analyticity of the elastic $b$-number form factor
$F$ defined as:
\beq
\la B(p')|\bar b \gamma^\mu b |B(b)\ra =(p+p')^\mu F(q^2)
\eeq
which is normalized as $F(0)=1$
in the large mass limit $M_B \simeq M_D$. Using the positivity of the
vector spectral function and a mapping in order to get a bound on the
slope of $F$
outside the physical cut, they obtained a rigorous but weak bound:
\beq
F'(vv'=1) \geq -6.
\eeq
Including the effects of the $\Upsilon$ states below $\bar BB$ thresholds
by assuming that the $\Upsilon \bar BB$ couplings are
of the order of 1, the
bound becomes stronger:
\beq
F'(vv'=1) \geq -1.5.
\eeq
Using QSSR, we can estimate the part
of these couplings entering in the elastic form factor.
We obtain the value of their sum \cite{SN7}:
\beq
\sum g_{\Upsilon \bar BB} \simeq 0.34 \pm 0.02.
\eeq
In order to be conservative, we have considered the previous estimate
within a factor 3 larger. We thus obtained the improved bound
\beq
F'(vv'=1) \geq -1.34,
\eeq
but the gain is not much, compared with the previous one. Using the
relation of the form factor with the slope of the Isgur--Wise function,
which differs by -$16/75 \log \alpha_s (M_b)$ \cite{FALK},
one can deduce the final bound:
\beq
\zeta'(1) \geq -1.04.
\eeq
However, one can also use the QSSR expression of the Isgur--Wise function
from vertex sum rules \cite{NEU} in order to extract
the slope $analytically$. The $physical$ IW function reads:
\bea
\zeta_{phys}(y\equiv vv')= \ga \frac{2}{1+y} \dr ^2 \Bigg \{
1 +\frac{\alpha_s}{\pi}
f(y) -\la \bar dd \ra \tau^3 g(y) + \\ \nnb
\la \alpha_sG^2 \ra \tau^4 h(y)
+g\la \bar dGd \ra \tau^5 k(y)+... \Bigg \},
\eea
where $\tau$ is the Laplace sum rule variable
and $f,~ h$ and $k$ are analytic functions of $y$. From this expression, one
can derive the analytic form of the slope \cite{SN7}:
\beq
\zeta'_{phys}(y=1) \simeq -1 + \delta_{pert} + \delta_{NP},
\eeq
where at the $\tau$-stability region:
$
\delta_{pert} \simeq -\delta_{NP} \simeq -0.04,
$
which shows the near-cancellation of the non-leading corrections.
Adding a generous $50 \%$
error of 0.02 for the correction terms, we finally deduce:
\beq
  \zeta'_{phys}(y=1) \simeq -1 \pm 0.02.
\eeq
Using this result in different existing model-parametrizations,
we deduce the value of the mixing angle:
\beq
V_{cb} \simeq \ga\frac{1.48\mbox{ps}}{\tau_b}\dr^{1/2}\times
(37.3 \pm 1.2 \pm 1.4)\times 10^{-3},
\eeq
where the first error comes from the data and the second one from the
model dependence.

\nin
 Let us now discuss, the effects due to the
$1/M$ correction. It has been argued recently (but the situation
is still controversial \cite{VAIN}), that the $1/M^2$ effect
can lower the Isgur--Wise function to a value  $0.89 \pm 0.03$
 at $y=1$ \cite{VAIN},
 such
that the extracted value of $V_{cb}$ using an extrapolation
until this particular point will
also increase by 11$\%$. However, the data from different groups near
this point are very inaccurate and lead to an inaccurate
though model-independent result.
 Moreover, in order to see the real effect of the $1/M$ correction,
 one can combine this previous result at $y=1$ with the sum rule estimate
of the relevant form factor at $q^2=0$, which is about $0.53 \pm 0.09$
\cite{SN5}. With these two extremal boundary
conditions and using the linear
parametrization:
\beq
\zeta = \zeta_0 +\zeta'(y-1),
\eeq
we can deduce the
slope:
\beq
 \zeta' \simeq -( 0.72 \pm 0.2).
\eeq
 It indicates that the $1/M$
correction tends also to decrease $\zeta'$, which implies that, for
larger values of $y$ where the data are more accurate,
the increases of $V_{cb}$ is weaker (+ 3.7$\%$) than the
one at $y=1$, which leads to the $final$ estimate:
\beq
V_{cb} \simeq \ga\frac{1.48\mbox{ps}}{\tau_b}\dr^{1/2}\times
(38.8 \pm 1.2 \pm 1.5 \pm 1.5)\times 10^{-3},
\eeq
where the new last error has been induced by the error from the slope.
This result is more precise than the one obtained at $y=1$, while the
model-dependence only brings a relatively small error.
It also shows that the value from the exclusive channels
is lower than that from the inclusive one,
 which is largely
affected by the large uncertainty
in the mass definition which enters in its fifth power.
Previous results for the slope and for $V_{cb}$ are in good
agreement with the new CLEO data presented in this meeting.
\section{ Hybrid and $B_c$-mesons}
Let me conclude this talk by shortly discussing the masses of the
hybrid $\bar QGQ$ and the mass and decays of the $B_c$-mesons.
Hybrid mesons are interesting due to their exotic quantum numbers. Moreover,
it is not clear if these states are true resonances or instead they only
manifest as a wide continuum. The lowest $\bar cGc$ states appear to be a
$1^{+-}$ of a mass around 4.1 GeV \cite{SNB}.
The available sum-rule analysis of
the $1^{-+}$ state is not very conclusive due to the absence of stability
for this channel. However, the analysis indicates
that the spin one states are in the range 4.1--4.7 GeV. Their characteristic
decays should occur via the $\eta'$ U(1)-like particle produced together with a
$\psi$ or $\eta_c$. However, the phase-space
suppression can be quite important for these reactions.
The sum rule predicts
that the $0^{--},~0^{++}$ $\bar c Gc$ states are in the range 5-5.7 GeV, i.e
about 1 GeV above the spin 1.

\nin
We have estimated the $B_c$-meson mass and coupling by combining the results
from potential models and QSSR \cite{BC}. Potential models predict:
\bea
M_{B_c}= (6255 \pm 20) ~\mbox{MeV},~~~
M_{B^*_c}=(6330 \pm 20)~\mbox{MeV} \\  \nnb
M_{\Lambda(bcu)}= (6.93\pm 0.05)~\mbox{GeV},~~~
M_{\Omega(bcs)}=(7.00 \pm 0.05)~\mbox{GeV} \\ \nnb
M_{\Xi^*(ccu)}=(3.63 \pm 0.05)~\mbox{GeV},~~~
M_{\Xi^*(bbu)}=(10.21 \pm 0.05)~\mbox{GeV}
\eea
which are consistent with, but more precise than,
the sum-rule results. The decay constant of the $B_c$
meson is better determined from QSSR. The average of the sum rules with the
potential model results reads:
\beq
f_{B_c} \simeq (2.94 \pm 0.12)f_\pi ,
\eeq
which leads to the leptonic decay rate into $\tau \nu_\tau $ of
about $(3.0 \pm 0.4)\times (V_{cb}/0.037)^2 \times 10^{10}~ \mbox{s}^{-1}$.
We have also studied the semileptonic decay of the $B_c$ mesons
and the $q^2$-dependence of the form factors.
We found that in all cases, the QCD predictions
increase faster than the usual pole dominance ones.
The behaviour can be fitted with an effective pole mass of about 4.1--4.62 GeV
instead of the 6.33 GeV one expected from a pole model.
Basically, we also found that
the each exclusive channel has almost the same rate of about 1/3 of the
leptonic
one. Detections of these particles in the next $B$-factory machine will serve
as a stringent
test of the results from the potential models and sum rules
analysis.
\section{Conclusion}
We have shortly presented different results from QCD spectral sum rules
in the heavy-quark sector,
which are useful for further theoretical studies
and complement the results from
lattice calculations or/and heavy-quark symmetry.
For the experimental point of view,
QSSR predictions agree with available data
but they also lead to some new features,
 which need to be tested in forthcoming
experiments.

\noindent
\vfill \eject

\end{document}